\documentclass[
    ,final            
  ]
  {aipproc}

\layoutstyle{8x11double}

\usepackage{graphicx,latexsym,amssymb}

\begin{document}
\newcommand {\sax} {{\it Beppo}SAX }
\newcommand {\rosat} {{\it ROSAT }}
\newcommand {\rxte} {{\it RXTE} }
\newcommand {\swift} {{\it Swift} }
\newcommand {\chandra}{{\it Chandra} }
\newcommand {\hess} {{HESS }}
\newcommand {\rchisq} {$\chi_{\nu} ^{2}$} 
\newcommand {\chisq} {$\chi^{2}$}
\newcommand {\ergs}[1]{$\times10^{#1}$ ergs cm$^{-2}$ s$^{-1}$}
\newcommand {\e}[1]{$\;\times10^{#1}$}
\newcommand {\nufnu}{$\nu F_{\nu}$ }
\newcommand {\nw}{nW m$^{-2}$sr$^{-1}$}
\newcommand {\micron}{$\mu$m }
\newcommand {\cms}{cm$^{-2}$s$^{-1}$ }
\newcommand {\mjd}{MJD$^\star$}


\title{\vspace{-1.3cm} The new surprising behaviour of the two  \\
"prototype" blazars 
PKS\,2155-304 and 3C\,279}


\author{Luigi Costamante}{
  address={HEPL/KIPAC Stanford University, Stanford, CA 94305-4085, USA}
}

\author{Felix Aharonian}{
  address={Max-Planck-Institut f\"ur Kernphysik, Saupfercheckweg 1, 69117 Heidelberg, Germany}
}

\author{Rolf B\"uhler}{
  address={Max-Planck-Institut f\"ur Kernphysik, Saupfercheckweg 1, 69117 Heidelberg, Germany}
}
\author{Dmitry Khangulyan}{
  address={Max-Planck-Institut f\"ur Kernphysik, Saupfercheckweg 1, 69117 Heidelberg, Germany}
}

\author{\\ Anita Reimer}{
  address={HEPL/KIPAC Stanford University, Stanford, CA 94305-4085, USA}
}

\author{Olaf Reimer\vspace{-0.1cm}}{
  address={HEPL/KIPAC Stanford University, Stanford, CA 94305-4085, USA}
}

\begin{abstract}
Recent VHE observations have unveiled a surprising behaviour 
in two well-known blazars at opposite sides of the blazar sequence.
PKS\,2155-304  have shown for the first time in an HBL a large Compton dominance, 
high $\gamma$-ray luminosities and a cubic relation between X-ray and VHE fluxes.
3C\,279 is the first FSRQ detected at VHE. The high luminosity required
to overcome the significant absorption caused by the BLR emission cannot be easily 
reconciled with the historical and quasi-simultaneous SED properties.
Both cases shed a new light on the structure and ambient fields of blazars.
Contrary to previous claims, it is also shown that
3C\,279 --as any FSRQ-- cannot in general provide  robust constraints on the EBL.
\end{abstract}

\maketitle


\section{Introduction}
\vspace{-0.2cm}
At opposite sides in the blazar sequence \cite{gg98}, 
the well-known,  bright blazars PKS\,2155-304 and 3C\,279  
are prototypes of the two blazar classes of
high-energy peaked BL Lacs (HBL) and Flat spectrum radio quasars (FSRQ). 
%
PKS\,2155-304 is a classic HBL, characterized by a spectral energy distribution (SED)
peaking in the UV-X-ray and GeV-TeV bands, 
by the X-ray band dominated by synchrotron emission of high-energy electrons, 
by the absence (or very low intensity) of the Broad Line Region (BLR) emission 
(typical HBL upper limits at $\lesssim10^{41}$ erg/s), 
and by a Compton dominance $L_{\rm C}/L_{\rm S}\lesssim1$. 
So far HBL have been successfully explained with a pure and homogeneous 
synchrotron self-Compton  model (SSC, see e.g. \cite{maraschi92,gg98}).
3C\,279 instead is a classic  FSRQ, characterized by intense BLR emission  
($L_{\rm BLR}\simeq 2.4\times10^{44}$ erg/s \cite{pian}, 
though 
FSRQ can reach $10^{46}$ erg/s), 
by a low-energy-peaked SED  (with humps located at IR and MeV-GeV frequencies),
by the X-ray band dominated by the inverse Compton (IC) emission of low-energy electrons,
and by a very high Compton dominance (up to $L_{\rm C}/L_{\rm S}\sim100$). 
The redder SED and high Compton dominance have been successfully explained by Comptonization 
on the external BLR photons, which together with the higher synchrotron luminosity
yield stronger cooling for the electrons \cite{gg98,ggsequence2}.

Since several decades, these two blazars have been observed  extensively
in all wavebands, except at the highest energies. With the new generation 
of Cherenkov Telescopes, since 2002 the Very High Energy ($\gtrsim100$ GeV) band started 
to be explored with sufficient sensitivity. 
At first these two objects followed the expectations for their class and related 
emission scenarios.  But in 2006, both sources managed to surprise us with new, unprecedented 
behaviours,  showing for the first time properties which were previously seen only 
in the opposite class. 
These properties now highlight a more complex role of the ambient fields and 
external Compton process in both types of sources, and provide us with a new level of 
insights on the jet structure and emission mechanisms in blazars.

\begin{figure}
 \includegraphics[height=.21\textheight]{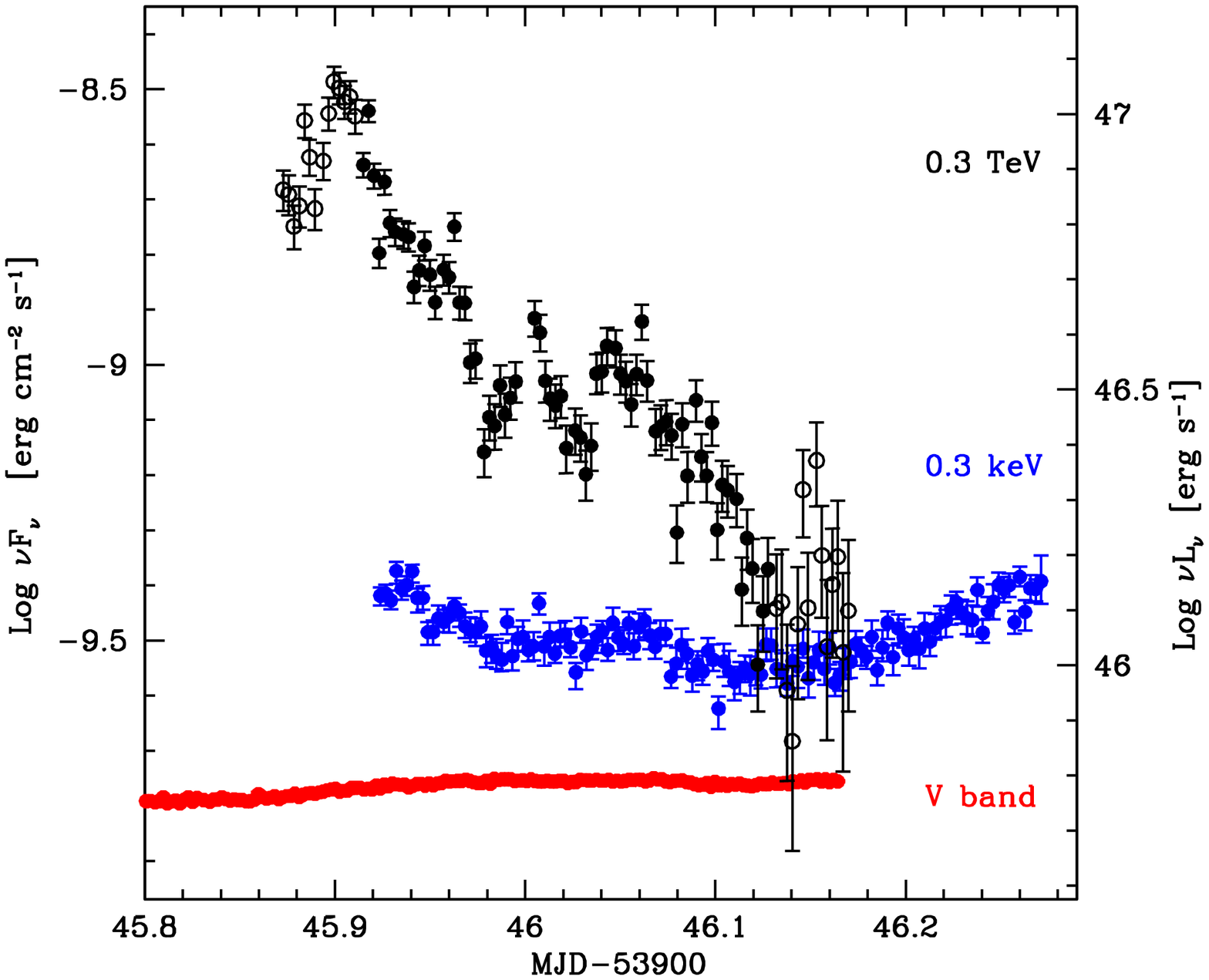}
 \includegraphics[height=.21\textheight]{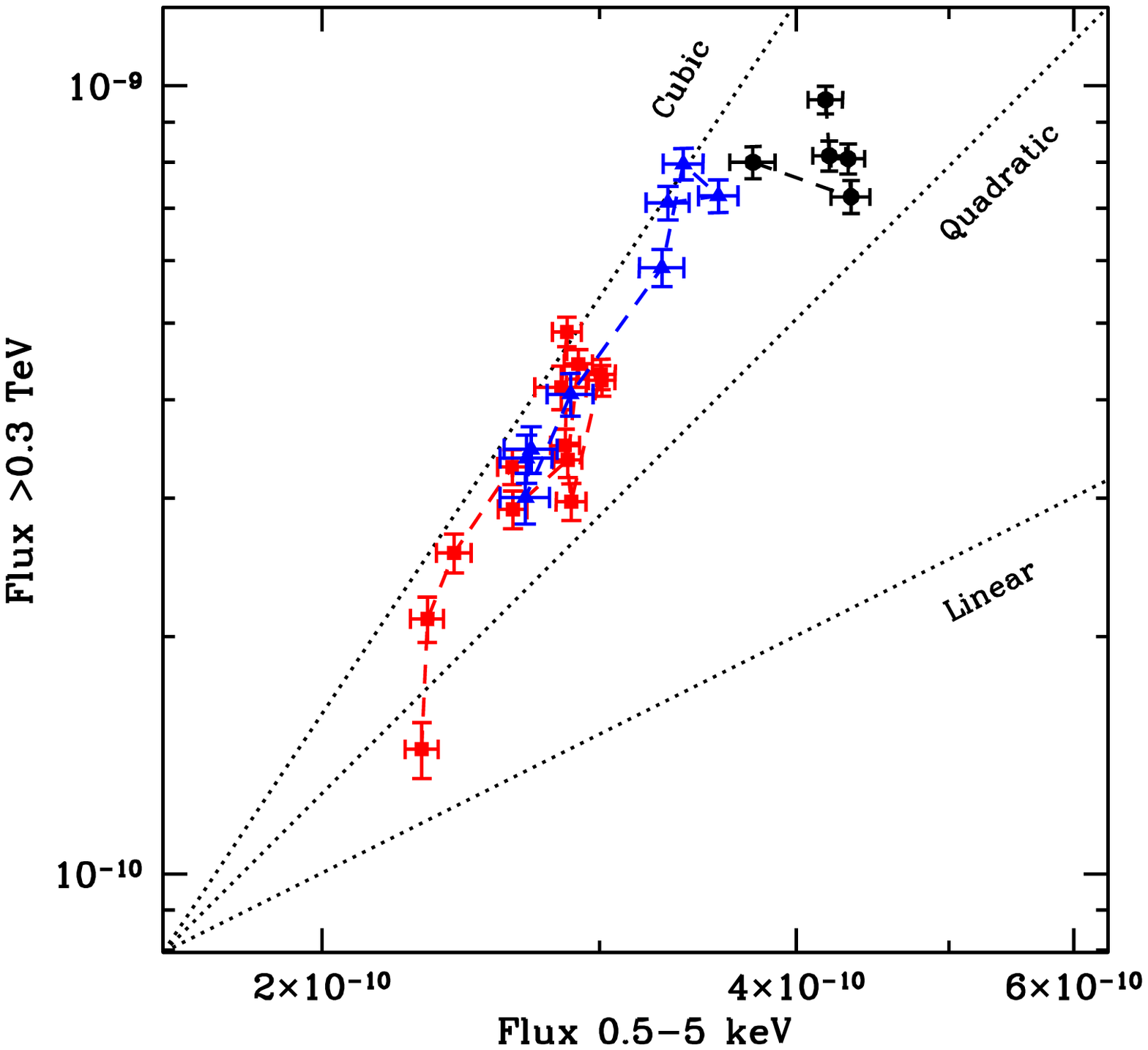}
 \includegraphics[height=.21\textheight]{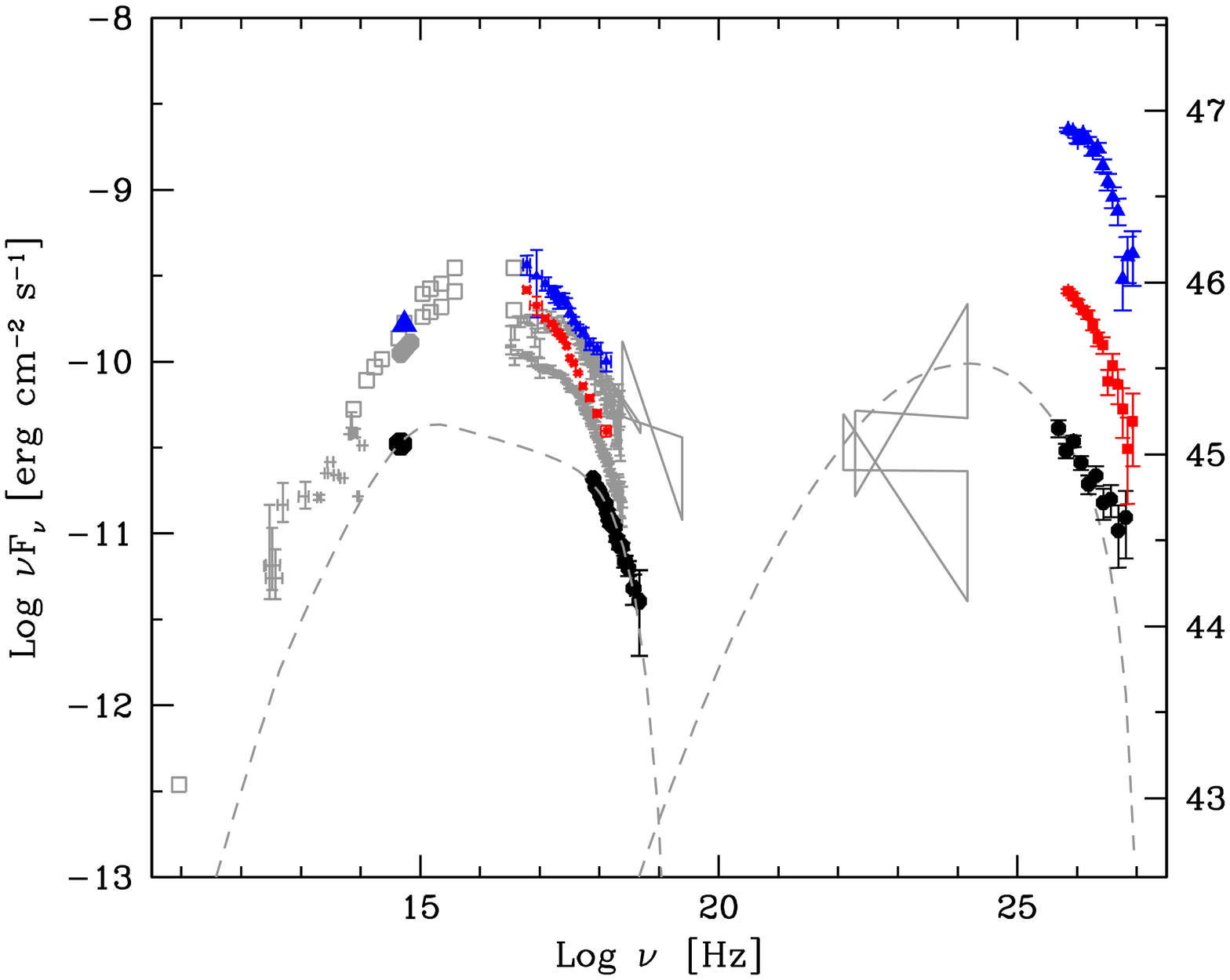}
  \caption{PKS\,2155-304 activity on July 29-30.  
  Left: lightcurve of the $\nu F_{\nu}$ flux at 0.3 TeV (EBL-corrected), 0.3 KeV and 5500\AA, in 4-min bins.
  Center: VHE vs X-ray flux in 7-14 minute bins, in a log-log plot and same units 
  (erg cm$^2$ s$^{-1}$). Right: SED of the two highest
  and lowest states in the simultaneous X-ray/TeV/optical window, together with historical data
  (details in \cite{costa2155}). \vspace*{-0.32cm} }
\end{figure}

\vspace{-0.4cm}
\section{The new PKS 2155-304: Compton-dominated HBL}
\vspace{-0.2cm}
In July-August 2006, PKS\,2155-304 entered a phase of  exceptional activity at VHE
monitored by HESS and other observatories (for a full overview, see \cite{lenein}). 
On the night of July 29-30, a simultaneous campaign with \hess, {\chandra} and the 
Bronberg optical observatory was performed, obtaining an unprecedented 6-8 hours
of continuous coverage in the three bands \cite{costa2155}.
Figure 1 shows the main results: the source showed  one major flare along the night 
in all three energy ranges, but with very different variability amplitudes.
For the first time, an HBL is characterized by a large Compton dominance  
(L$_{\rm C}$/L$_{\rm S}\sim$8-10) -- which evolves in few hours
to the more usual value of 1 --  and shows a {\it cubic} relation between 
VHE and X-ray flux variations, during a decaying phase.
The emission in the X-ray and VHE bands are highly correlated, 
both in flux and spectrum, while the optical ligthcurve 
shows a flare which starts simultaneously with the VHE flare, 
but remains constant afterwards.

In a typical one-zone SSC model, it is difficult to account for a cubic correlation between 
synchrotron and IC emissions above the respective SED peaks and during a decaying phase.
On one hand, the correlated variability (and flare on-set) does indicate that the emission 
in all bands likely originates from the same flaring event,  and possibly same emitting region.
On the other hand,  even accepting a large bulk-motion Lorentz factor
as implied by a one-zone analysis (e.g. as in \cite{tavecchio98}, yielding 
$\Gamma\gtrsim100$ and $B\lesssim0.005$ G) and to have the X-ray emission in the Thomson regime,
at most a quadratic relation can be explained \cite{katar05}.
A possibility is to invoke coincident, fast variations of the magnetic field B, 
but anti-correlated with the flux variations: namely B increases as the flux decreases.
This would enhance the cooling of electrons through the synchrotron channel,
further suppressing the VHE emission while at the same time ``keeping up"
(in part or totally) the X-ray synchrotron flux.
However, 
such possibility seems excluded by the optical data:
fast changes of B would lead to correlated variations of the synchrotron emission 
($\propto B^2$)  of the lower-energy electrons, which emit longward of the synchrotron peak
and have not yet cooled (with the aforementioned one-zone parameters). 
This is contrary to observations: the optical flux remains almost constant in the affected interval.

A more viable explanation seems to be the superposition of two
emitting zones. A steady one responsible for the usual ``persistent" SED of PKS\,2155-304
(peaking in the UV and with low VHE emission), and a second zone 
--more compact and with larger bulk motion-- responsible for the flaring activity.  
The X-ray (i.e. synchrotron) variations of this second component can thus be as large as 
the VHE ones, but are simply seen ``diluted" in the ``persistent" SED, 
while they are fully visible in the VHE band \cite{costa2155}.
A linear relation between VHE and X-ray fluxes means  however
that the SED of this new component must have a high Compton dominance ($\approx$20) 
constant in time during the flare evolution. 
Such behaviour would point towards an origin of the 
$\gamma$-ray peak by external Compton rather than a pure SSC mechanism.
Indeed, this is expected in scenarios with a strong radiative interplay between 
different parts of the jet\cite{ggrapid,markos03}.
%
\indent A two-zone scenario is common to explain major flares in blazars.
However, so far all previous events that got extensive VHE sampling
have shown flaring components  which were synchrotron-dominated, 
often leading to  dramatic shifts of the overall SED peak, as seen in 
Mkn\,501 or 1ES\,1959+650.
The novelty of this event is that the bulk of the luminosity of
an otherwise comparable flare ($\sim10$ times the average source apparent luminosity) 
is now emitted in the Compton channel instead of the synchrotron channel.
A bimodality seems thus to emerge in the mode of flaring for HBL: 
either synchrotron or Compton-dominated. More observations are needed to
assess if this is only a rare event or a new common feature of the HBL class.

\begin{figure}[t]
  \includegraphics[height=.25\textheight]{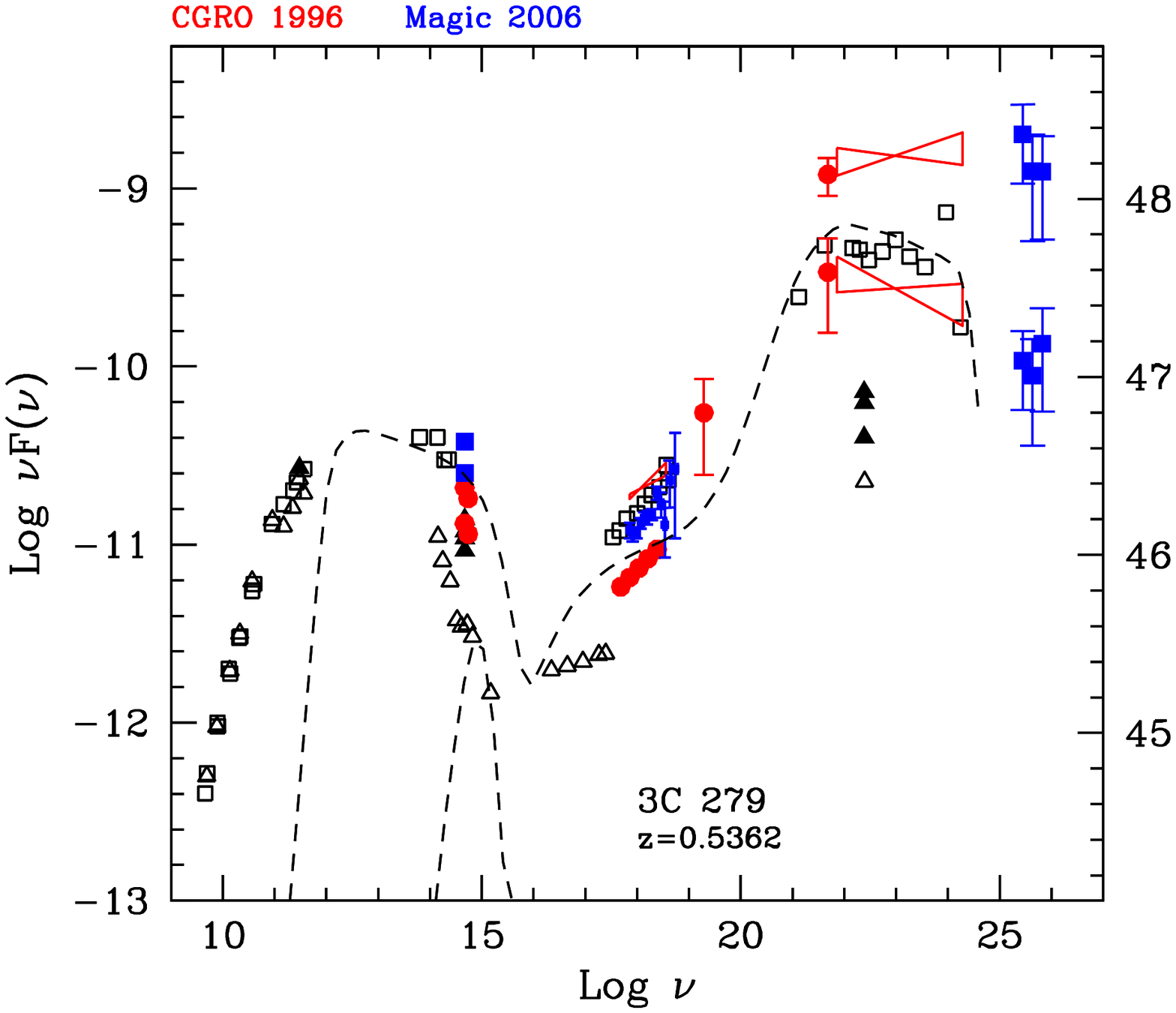}
  \includegraphics[height=.30\textheight]{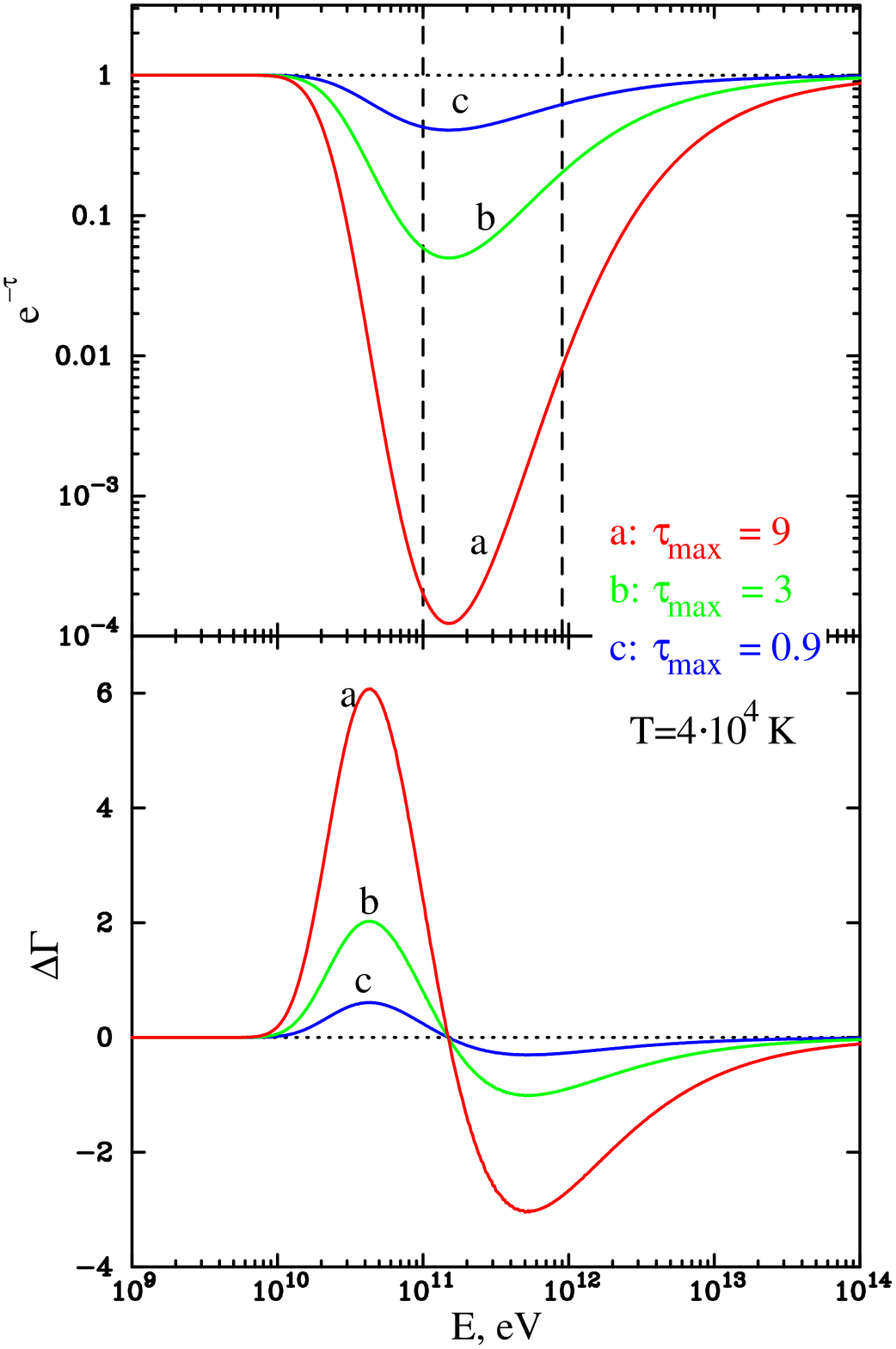}
  \includegraphics[height=.21\textheight]{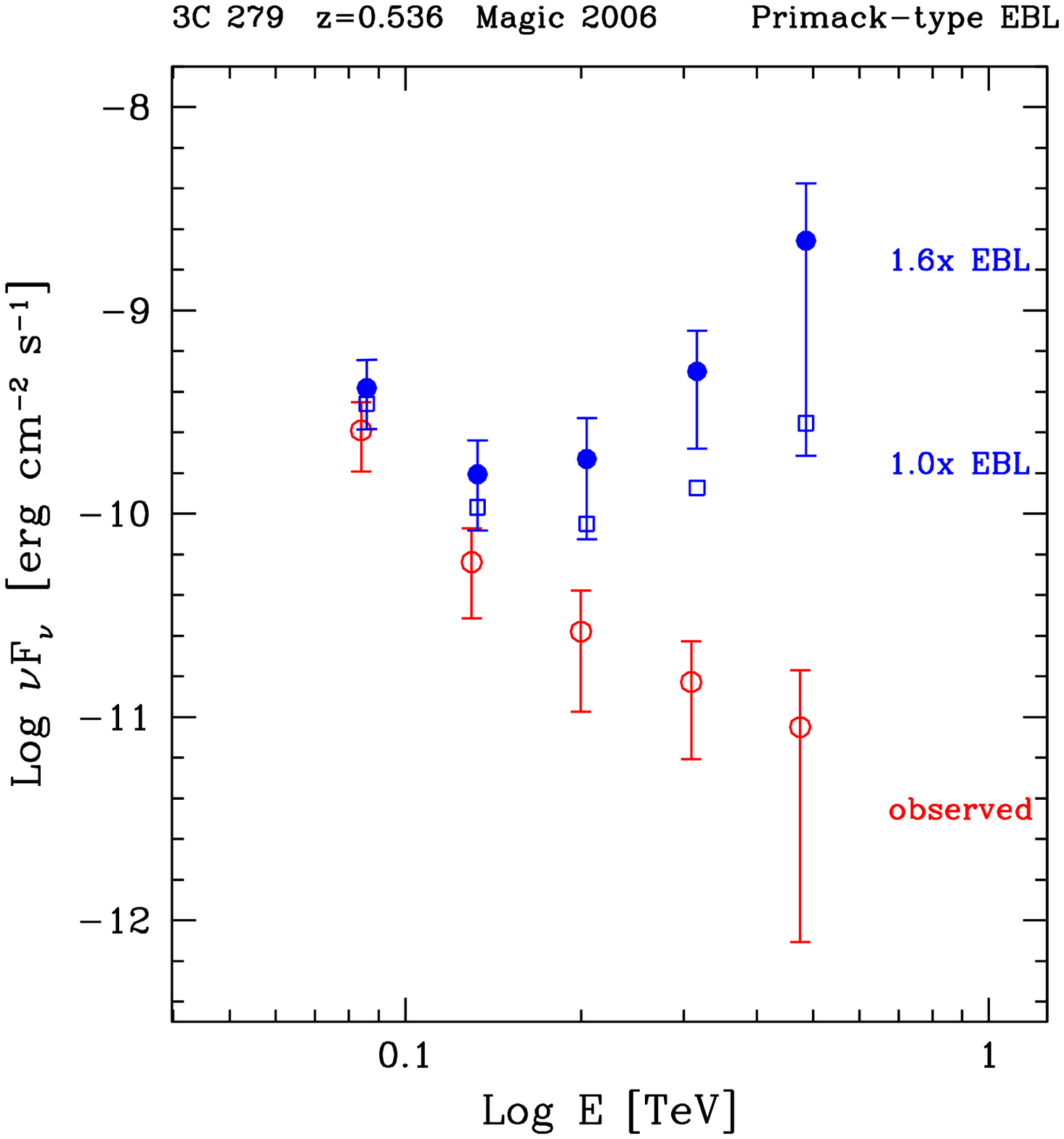}
  \caption{Left: historical SED of 3C\,279 \cite{gg98} with in red the CGRO high states in 1996 
   \cite{wehrle}. In blue, the simultaneous optical and VHE fluxes
   (from \cite{magic}) and the quasi-simultaneous RXTE spectrum. Given the large uncertainty on the
   VHE spectrum, only the 100-300 GeV data are shown, for flux reference. Center, upper panel:
   attenuation factor $e^{-\tau}$ for absorption on BLR field approximated as 
   blackbody with temperature T$=4\times10^4$K, for three values of $\tau_{\rm BLR}$. 
   Vertical lines mark the rest-frame energy range corresponding to the
   MAGIC spectrum. Lower panel: variation of the local photon index of the 
   emerging spectrum with respect to the incident spectrum (details in \cite{fkc08}).
   Right: the whole MAGIC spectrum \cite{magic} corrected for EBL absorption with an EBL shape  
   as in \cite{primack} (including galaxy evolution; open squares), and rescaled to 
   the level of the upper limits in \cite{nature_ebl}
   (filled circles). \vspace*{-0.4cm}}
\end{figure}

\vspace{-0.35cm}
\section{The puzzle of 3C\,279: first FSRQ detected at VHE}
\vspace{-0.2cm}
The MAGIC observation  
of 3C\,279 ($z=0.536$) in Feb. 2006 \cite{magic}
marks the first detection at VHE of both a FSRQ and a source at $z\gtrsim0.5$.
Contrary to what generally believed, however,
the redshift is {\bf not} the surprise and main aspect of this discovery. 
In past years, the VHE spectra of several HBL 
have already indicated that the Universe is more transparent to $\gamma$-rays than
previously thought, with an EBL density close to the lower limits given by galaxy counts 
\cite{nature_ebl, hess0347}.
As a consequence, it was immediately realized that sources as far as $z=0.5-0.6$ 
can indeed be detectable around 100-300 GeV with the current-generation instruments, 
without requiring extreme flux states or modifications in fundamental laws of physics \cite{costa_barca}.
Specific targets were also proposed, and more detections are expected. 

The main suprise is represented by the source being a FSRQ, namely an object with 
{\it a)} intense BLR emission ($L_{\rm BLR}\simeq 2.4\times10^{44}$ erg/s \cite{pian}),
and {\it b)} a low-energy peaked  SED (Fig. 2), implying few high-energy electrons available 
to sustain a strong VHE emission.
This represents a problem because a high VHE luminosity is instead required to
overcome both the extragalactic EBL absorption (high due to the large distance)
and the strong {\it internal} absorption caused by $\gamma$-$\gamma$ collisions with
the BLR photons.  
The BLR emission is peaked in the UV range (typically $\sim$9 eV, rest-frame, 
around the most prominent lines HeII, Ly$\alpha$ and CIV). 
This is precisely the energy range where the $\gamma$-$\gamma \rightarrow e^+e^-$ 
cross section is maximum for 100-200 GeV photons.
Adopting the line luminosity observed in 3C\,279, a typical covering factor $\sim$10\% 
and the observational relation between BLR size and disk luminosity 
($R_{\rm BLR}\propto L_{disk}^{1/2}$ \cite{bentz2006,kaspi2007}),
the resulting BLR energy density is of the order of
$U_{\rm BLR}\simeq 10^{-2}$ erg cm$^{-3}$ (ranging bewteen 0.7 and 10) 
within a radius of $R_{\rm BLR}=1-4\times10^{17}$ cm.
This implies a maximal optical depth $\tau$ at 100-200 GeV (rest-frame)  which can be as high as
$\tau_{{\rm BLR},max}\propto 9 \,l_{17}$, where $l_{17}$ is the $\gamma$-ray 
photon path inside the BLR, in units of $10^{17}$ cm.

Figure 2 (left) shows the SED of 3C\,279 with the VHE fluxes of the MAGIC detection
corrected one time for EBL absorption  (using \cite{primack}) 
and a second time with $\tau_{{\rm BLR,max}}=3$.  The resulting SED shape suggests
that the true initial VHE luminosity can likely be close or above the highest EGRET fluxes 
($3\times 10^{48}$ erg/s \cite{wehrle}), if the MAGIC flare took place
inside the BLR. In such case, the SED suggests that the EGRET flat spectrum observed in high state
could extend up to 100-300 GeV with no sign of cutoff.  
To produce VHE photons  by IC, very high energy electrons are required by energy-conservation law 
($\gamma\geq 6*10^5/\delta$), which emit by synchrotron in the UV-X-ray range 
for the typical parameters used in blazars ($h\nu_{sync}\geq 4*B/\delta$ keV).
Yet, the quasi-simultenous RXTE spectrum (taken 0.3 and 1 days apart from the MAGIC pointings, during
routine monitoring) is hard (photon index $\Gamma_X=1.66\pm0.11$),  
as typical for this source and FSRQ in general, and usually interpreted as the emerging
of the IC emission.  
This X-ray spectrum represents a strict upper limit on the synchrotron luminosity of 
TeV electrons,  unless one adopts the different and unconventional view 
that such X-ray spectrum corresponds actually to synchrotron emission of a second, 
``extreme BL Lac"-like component in the jet of 3C\,279.
In the leptonic scenarios, the hard X-ray spectrum provides also a limit
for the $\gamma$-ray luminosity absorbed in the BLR,
since the resulting pairs would reprocess the VHE power into the X-ray band
through IC on the BLR UV photons, leading to much softer X-ray spectra \cite{ggmadau96}.
Upcoming GLAST observations will be crucial to solve all these issues.
If confirmed in strictly simultaneous observations with Cherenkov telescopes, 
the 3C\,279 SED might become more easily explained with hadronic rather 
than leptonic scenarios, like the proton-synchrotron model \cite{felix2000,anita01} 
(see \cite{bott} for a detailed discussion on modelling and implications).

The energetic requirements are less if the VHE flare took place outside the BLR,
but in such case the external Compton process cannot use the BLR photons as target field.
Also the near/mid-IR photons from hot dust\cite{sikora_irc} cannot be used in this case, since
the implied energy densities would typically suppress most of the radiation approaching 1 TeV.
(unless assuming extreme values of bulk motion).
The huge Compton dominance and low-energy peaked SED have to be explained otherwise, 
e.g. using seed photons from different parts of the jet
(as in the spine-layer or decelerated jet scenarios, \cite{ggspinelayer,markos03}), 
or as a purely SSC flaring episode. In the latter case, however,  larger X-ray fluxes
and a more "HBL-like" SED are expected, which should be revealed
by strictly simultaneous observations.



\vspace{-0.2cm}
\subsection{Not any blazar is suited for EBL studies}
\vspace{-0.1cm}
\noindent In the MAGIC paper \cite{magic}, the (admittedly poorly determined) 
VHE spectrum of 3C\,279 has been used to derive constraints on the EBL 
as if this source was an HBL.  However, the presence of the BLR target field 
in this object --as in any FSRQ--
cannot be  neglected, since it can strongly affect the spectrum emerging from the source.
Such influence has been discussed since the COS-B discovery of 3C\,273 in 1979 
\cite{mcbreen}, but its impact on EBL limits has been recently 
recognized by \cite{anita} for the redshift dependence  
and by \cite{fkc08} for the spectral hardening
(see also \cite{sitarek,liu}). It has been shown that absorption
on a narrow-banded target field leads to the formation of $\gamma$-ray
spectra of almost arbitrary hardness,
irrespective of the primary spectrum emitted by the source \cite{fkc08}.
This effect undermines any conclusion on EBL limits that can be drawn
from the hardness alone of $\gamma$-ray spectra in FSRQ.
The spectral hardening depends primarily on the intensity of the target field:
the effect is shown in Fig. 2 (center), for a black-body target spectrum 
with $T=4\times10^4 K$  (generally adopted as a good approximation 
of the  BLR radiation field \cite{ggsequence2}) 
and for three different value of the optical depth $\tau$.
The only limit on the achievable hardness is represented by the higher
apparent luminosity required to the source. 
Adopting for reference the largest EGRET fluxes measured in FSRQ, and specifically 3C\,279,
the MAGIC flux around 100 GeV ($\simeq10^{47}$ erg/s) allows a combined
internal$+$intergalactic optical depth as large as $\tau_{\rm BLR}+\tau_{\rm EBL}\sim5$.
Since $\tau_{\rm EBL}\simeq0.4-2$, $\tau_{\rm BLR}$ can be realistically
in the range 4.5-3, implying a possible severe hardening (see Fig.2).

This hardening is avoided if either the emitting zone is outside the BLR,
or the effective BLR spectrum is much more broad-banded than considered.
The latter case has been recently argued by \cite{tavecchiobis}, who showed that,
by assuming a more detailed model for the AGN disk emission, the BLR
spectrum is broader towards the optical frequencies, resulting in an optical depth 
almost constant with energy in the MAGIC passband.  
However, this model introduces a discrepancy in the normalization
of the BLR Size-Luminosity relation derived in the UV (1350\AA \cite{kaspi2007})
and Optical bands (5100\AA \cite{bentz2006}), 
precisely because too much broad-banded. The BLR size results very different
depending on the chosen band, for the same disk/BLR luminosity. 
A narrower field instead, close to a black body between the two wavebands, 
yields similar values for the BLR size, and is thus supported by those data.
Moreover, if one is forced to consider the MAGIC spectral shape accurate 
enough to derive limits on the EBL despite its large statistical uncertainty 
($>\pm1$ at 1$\sigma$, once deabsorbed), 
the EBL-corrected spectrum shows (with any EBL model) an  odd concave shape  
as indeed expected
from internal absorption on a narrow-band field 
peaking at the BLR line energies (compare right with center panels in Fig. 2). 

In conclusion, the important point is that the actual BLR spectrum as seen by the gamma-rays 
is highly uncertain, with observational evidence both pro and against the  narrow-band hypothesis.
As long as the hardening effect cannot be excluded, therefore, {\it no robust conclusion
can be derived on the EBL from 3C\,279 alone, or from any other object with strong emission lines},
contrary to what claimed in \cite{magic,tavecchiobis}.
Only those objects for which the emission zone is established beyond the BLR 
(e.g. without the cut-off in the 10-100 GeV range from BLR absorption) 
can possibly qualify\cite{canonical}. This problem concerns also some BL Lacs as well as FSRQ, 
since the relevant factors are the line luminosity and the (uncertain) location of the emitting region 
inside the BLR,  not the equivalent width  (which  defines the BL Lac class); 
in fact some LBL have shown line luminosities comparable with FSRQ \cite{scarpa97}.
The MAGIC result therefore, while neither robust nor innovative concerning EBL constraints,
sheds a new light on the structure and emission mechanisms of blazars.

\vspace{0.2cm}
\noindent {\footnotesize {\bf ACKNOWLEDGMENTS}:  LC thanks the HESS Collaboration for 
endorsing and supporting
the Chandra-HESS campaign, and F. Tavecchio for insightful discussions.}

\bibliographystyle{aipproc}   

\vspace{-0.6cm}
\bibliography{costamante}

\end{document}